\documentclass{aa}  
\DeclareUnicodeCharacter{2212}{-}
\usepackage{orcidlink}
\usepackage{soul}
\usepackage{graphicx}
\newrobustcmd{\B}{\bfseries}
\DeclareUnicodeCharacter{2064}{+}
\usepackage{txfonts}
\usepackage{hyperref}
\usepackage{lineno}
\begin{document}

\title{\textcolor{black}{Exploring Coronal Abundances of M~Dwarfs\\ at Moderate Activity Level}
}

   \author{J. J. Chebly\orcidlink{0000-0003-0695-6487}
          \inst{1,2,3}
          \and
          K. Poppenh\"ager\orcidlink{0000-0001-5052-3473}\inst{1,2}
          \and
          J. D. Alvarado-G\'omez\orcidlink{0000-0003-1231-2194}\inst{1}
          \and B.~\textcolor{black}{E}.~Wood\orcidlink{0000-0002-4998-0893}\inst{4}}

   \institute{Leibniz Institute for Astrophysics, An der Sternwarte 16, 14482, Potsdam, Germany\\
              \email{\textcolor{black}{judy.chebly@cea.fr}}
         \and
              University of Potsdam, Institute of Physics and Astronomy, Potsdam-Golm, 14476, Germany
         \and
               \textcolor{black}{Université Paris Cité, Université Paris-Saclay, CEA, CNRS, AIM, F-91191, Gif-sur-Yvette, France}
            \and
            Naval Research Laboratory, Space Science Division, Washington, DC 20375, USA\\}

   \date{Received April 02, 2024; accepted February 2, 2025}

  \abstract
   {Main sequence stars of spectral types F, G, and K with low to moderate activity levels exhibit a recognizable pattern known as the first ionization potential effect~(FIP~effect), where elements with lower first ionization potentials are more abundant in the stellar corona than in the photosphere. In contrast, high activity main sequence stars such as AB Dor~(K0), active binaries~along with M~dwarfs exhibit an inverse pattern, known as iFIP. }
   {We aim to determine whether or not the iFIP pattern persists in moderate-activity M~dwarfs.}
  {We used XMM-Newton to observe the moderately active M dwarf HD~223889 that has an X-ray surface flux of $\log F_{\rm X, surf}$~=~5.26, the lowest for an M~dwarf studied so far for coronal abundance patterns. We used low-resolution CCD spectra of the star to  calculate the strength of the FIP~effect quantified by the FIP bias~($F_{\rm bias}$) to assess the persistence of iFIP~effect in M~dwarfs.}
  {Our findings reveal an iFIP effect similar to that of another moderately active binary star GJ 338 AB, with a comparable error margin.~The results hint at a possible plateau in the $T_{\rm eff}$-$F_{\rm bias}$ diagram for moderately active M~dwarfs. }
   {Targeting stars with low coronal activity with coronal temperature between 2\,MK and 4\,MK is essential for refining our understanding of (i)FIP patterns and their causes. }

   \keywords{Stars: magnetic field, Stars: abundances, Stars: activity, Stars: low-mass, X-rays: stars}

   \maketitle
%

\section{Introduction}
\label{sect:intro}

Scientists have observed a significant difference in the composition of the solar corona and wind compared to the photosphere. In the solar corona, the elemental abundance ratio changes by a factor of 3 with a typical variation between 2 and 5 relative to the photosphere \citep{1963ApJ...137..945P, 1985ApJS...57..173M, 1992PhyS...46..202F}. The elements with enhanced abundances are characterized by a low \textbf{f}irst \textbf{i}onization \textbf{p}otential~(FIP) and include Al, Mg, Si, Ca, and Fe~($< 10$\,eV). This phenomenon has been detected by various means, including remote sensing techniques such as spectroscopic measurements and in situ observations as documented by \cite{2000PhyS...61..222F} and \cite{2000JGR...10527217V}. In contrast, elements with high FIP~values ($\geqslant$~$10$\, eV), such as C, N, O, Ne, and Ar, have coronal abundances similar to their photospheric values. This discrepancy in abundance is known as the FIP~effect.

The FIP~effect has been observed not only in our Sun but also in stars with solar-like properties, as demonstrated in several studies~\cite{1994AAS...185.4503D, 1999ApJ...516..324L, 2002ASPC..277..497G,2003csss...12..313R,Robrade2009}.\textcolor{black}{~In contrast, high activity level stars particularly in M~dwarfs and active binaries, are strongly associated with an inverse FIP~(iFIP) effect~\citep{Liefke2008}.~Younger, more active stars, such as AB~Dor~(K0) and especially early-G dwarfs, tend to exhibit inverse or no FIP~effect, whereas older and less active stars show a solar-like~FIP effect~\citep{Raassen2003,Telleschi2005}.}

In a subsample of stars, consisting of stars on the main sequence and having low to moderate X-ray luminosities of $\log~L_{\rm X} \leq 29$, a correlation between spectral type and coronal abundances has been demonstrated: M~dwarfs generally exhibit an inverse FIP effect, which transitions to a neutral FIP effect in mid-K~dwarfs, and ultimately to a solar-like FIP effect in early G~dwarfs~\citep{2010ApJ...717.1279W, 2012ApJ...753...76W}. M~dwarfs exhibiting a FIP effect instead of an iFIP effect have not been found so far.

The model proposed by \cite{2004ApJ...614.1063L} provides the most comprehensive explanation for the observed FIP and iFIP~effect in stars to date. This model focuses on the ponderomotive force generated by Alfvén waves, which leads to the separation of ions and neutrons within the chromosphere of the Sun and other stars.

The FIP~effect is explained through resonant Alfv\'en waves traveling along coronal loops, as shown in the studies by (\citealt{2015ApJ...805..102L, 2021ApJ...909...17L}). The transmission of waves between the chromosphere and the corona with energy fluxes for coronal heating in the range of $\mathrm{10^6}$ to $\mathrm{10^7}$\,  erg $\mathrm{cm^{- 2}}$ $\mathrm{s^{-1}}$ affects the abundance of elements in the upper chromosphere based on their ionization states. Elements with low-FIP, which are predominantly ionized in the chromosphere, experience a significant increase in abundance as they ascend into the corona. The high FIP~elements on the other hand, which are mainly neutral in the chromosphere, do not appear to be affected by these wave-induced processes.

\begin{figure}[hbt!] 
     \centering
         \includegraphics[width=\columnwidth]{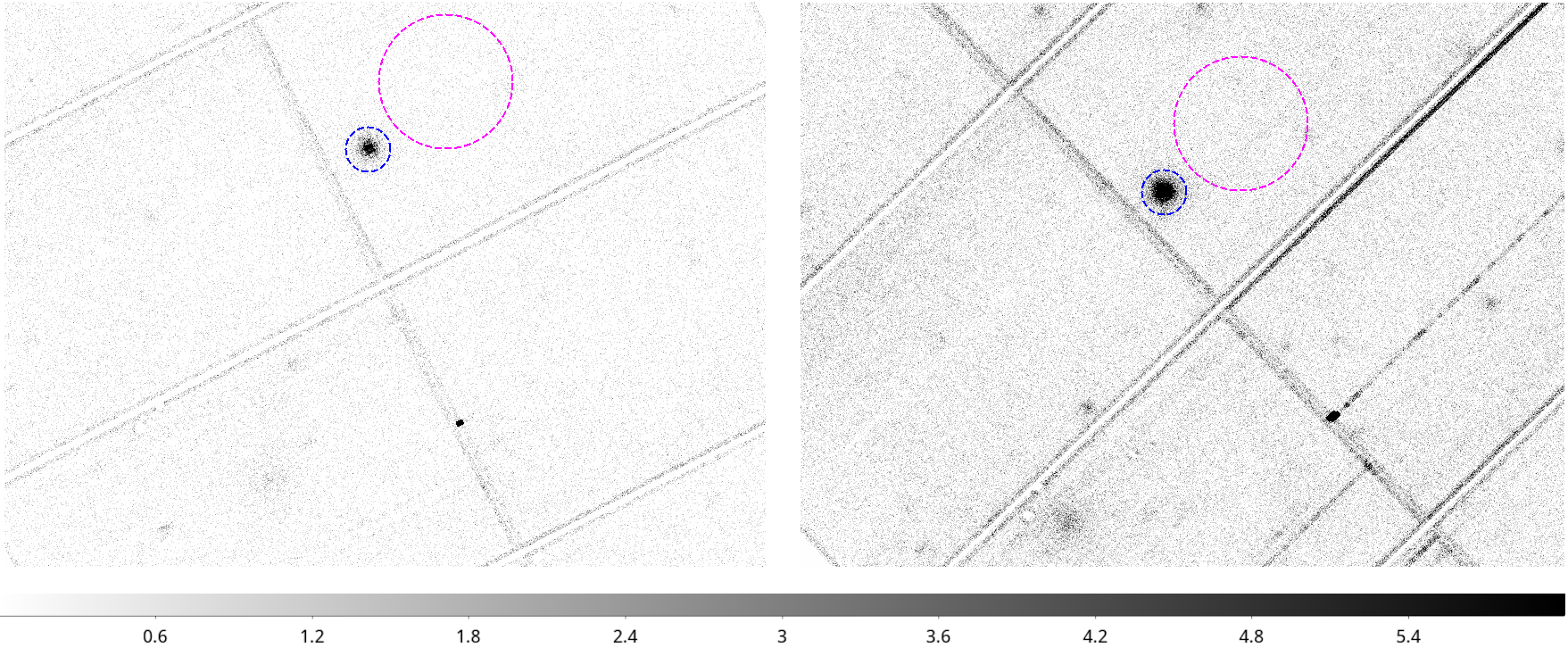}

     \caption{X-ray image from PN-detector of HD~223889 taken with the \textit{XMM-Newton} telescope in the $0.2$–$2$\,keV energy band. The left and right panel shows respectively, \textit{observation~1} and \textit{2} (see Table~\ref{tab:datainfo}). HD~223889 is marked by a blue dashed circle with a radius of 20~arcsec, while the background is represented by a magenta dashed-circle with a radius of 60~arcsec.}
\label{fig:X-ray}
\end{figure}

In contrast, the iFIP~effect is thought to be driven by the ability of upward-propagating p-modes or magneto-acoustic waves (waves driven by magnetic pressure) to undergo reflection or refraction back into the chromosphere. In the latter, the pressure of the magnetic field exceeds the thermal effects of charged particle motion, plasma $\beta < 1$~(\citealt{2019ApJ...875...35B, 2020ApJ...894...35B, 2021ApJ...909...17L}). 
In addition, iFIP is more likely to occur under conditions of restricted magnetic field expansion through the chromosphere (\citealt{2019ApJ...875...35B,2021ApJ...909...17L}). 
This is observed, for example, in stars with high filling factors\footnote{Proportion of the stellar surface covered by active regions.}. 

The filling factor of cool main sequence stars estimated from Zeeman broadening correlates with Rossby number~\footnote{Defined as $R_{\rm o} =$~rotation period/ convective turnover time} in a similar way to the activity-rotation relation~\citep{2011ApJ...741...54C, 2012LRSP....9....1R}. This means that more active stars have larger estimated filling factors. This is consistent with observations of M~dwarfs~(e.g. \citealt{2009ARA&A..47..333D, 2009A&A...496..787R}).
If this interpretation is correct, then stars with low magnetic activity and therefore low filling factors should display a FIP~effect, while stars with high filling factors display an iFIP~effect. 

It has also been observed that the transition from the FIP to the iFIP regime correlates with the stellar mass, especially when excluding extremely active stars characterized by a {$L_{\rm X}$ of $10^{29}$\,$\rm erg~s^{-1}$, as shown in~\cite{2010ApJ...717.1279W, 2012ApJ...753...76W, 2022A&A...659A...3S}. The pattern also decreases as we progress to the later spectral types, eventually reaching a null effect around K5, with an iFIP~effect for M~stars.~\textcolor{black}{Although the spectral type of a star has an influence on the resulting~FIP~effect, it is not the only determinant of this pattern. This becomes clear when we consider the case of Proxima Centauri~(M5V), which exhibits a mild iFIP. According to the studies of \cite{2004A&ARv..12...71G}, this effect is comparable to that observed in an M0V star such as GJ 338 AB, as documented by \cite{2012ApJ...753...76W}. If spectral type was the only determining factor, we would not expect such a similarity. Therefore, we must conclude that other factors might also play an important role in shaping the FIP~pattern.}

It is therefore currently unclear whether M~dwarfs with a moderate activity level~($27<\log~L_{\rm X} < 28$) can have low enough filling factors to display a FIP~pattern, instead of an iFIP pattern, in their coronal abundances. The lowest activity M dwarfs studied for coronal abundances so far are the two stars in the moderately active wide binary system GJ~338~AB, which consists of two M0~dwarfs located at a distance of about 5~pc from the Sun. The stellar system was the subject of a Chandra-LETGS observation performed by \cite{2012ApJ...753...76W}. The study showed the presence of a mild iFIP pattern in this binary star; however, the uncertainties in the measurements were large and almost encompassed the abundances in the solar photosphere. 

In this study, we investigate whether or not the iFIP pattern persists in low-to-moderately activity M~dwarfs by examining a critical coronal temperature range. Our study uses \textit{XMM-Newton} observations to investigate the coronal abundances of the nearby M~dwarf HD~223889, which is one of the lowest active M~dwarf suitable for studying the iFIP~effect.
The paper is structured as follows: Section \ref{sect:Obsv and analysis} gives an overview of the observations and the methods used for data analysis. Section \ref{sect: results} outlines the results of this study. Section \ref{sect: discussion} provides a detailed discussion and comparative analysis based on the findings from previous observations. The concluding remarks can be found in section \ref{sect:conclusion}.

\section{Observations and data analysis} 
\label{sect:Obsv and analysis}
The star HD~223889~(HIP~117828) is ~\textcolor{black}{of spectral type M2.5V-M3V, respectively \citealt{Maldonado2020,Gaidos2014}}.
The star is at a distance of $10.1$\,pc from the Sun~\textcolor{black}{(more details on the star can be found in table~\ref{star_properties}).~The stellar parameters are taken from~\cite{2019AJ....158..138S}, and from Gaia~DR3~\citep{2021A&A...649A...6G}.~When contextualizing our results alongside other cool stars, we chose to use the effective temperature derived from Gaia~DR3, as it offers a smaller margin of error, providing greater precision in our comparisons.} 

This star was observed twice with XMM-Newton in 2020 and again in 2022. HD~223889 is a good target for this study because although it has moderate coronal activity~$27<\log L_{\rm X} <28$, it has an estimated mean coronal temperature higher than 2\,MK~(determined from 2020 observation). At a coronal temperature of more than 2\,MK, iron emission with XMM-Newton can actually be observed. 

\begin{table}[htp]
    \centering
    \caption{HD~223889 intrinsic properties taken from \cite{2019AJ....158..138S}, and Gaia~DR3~\citep{2021A&A...649A...6G} are listed respectively in column 2 and 3.}
    \begin{tabular}{lll}
        \hline
        \hline
        Parameter & \cite{2019AJ....158..138S} & Gaia DR3 \\ 
        \hline
        Mass ($M_{\odot}$) & $0.52\pm 0.02$ & \\ 
        Radius ($R_{\odot}$) & $0.52 \pm 0.02$ & \\ 
        $T_{\text{eff}}$ (K) & $3517 {\pm 157}$ & $3254_{-12}^{+157}$\\ 
        $\log \rm g$ & $4.7 \pm 0.26$ & $4.1^{\rm + 0.4}_{\rm - 0.04}$ \\
        \\
        \hline
    \end{tabular}
    \label{star_properties}
\end{table}

The details of the observations are listed in table~\ref{tab:datainfo}. To distinguish between the two observations, we will refer to the 2020~observation as the "observation~1" and the 2022~observation as the "observation~2".
Observation~2 extended over 100~ks, whereas observation~1 extended only over 28~ks. Both observations used the EPIC\footnote{EPIC: European Photon Imaging Camera} and RGS\footnote{RGS: Reflection Grating Spectrometer} instruments of the XMM-Newton observatory to obtain both CCD and RGS spectra.
Notably, the EPIC observations (MOS1, MOS2, and PN)\footnote{MOS: Metal Oxide Semi-conductor}\footnote{PN: Proton-Noise camera} used the same filter configuration as observation~1, ensuring the consistency in the observations. 

\begin{figure*}[hbt!] 
     \centering
       \includegraphics[width=\textwidth]{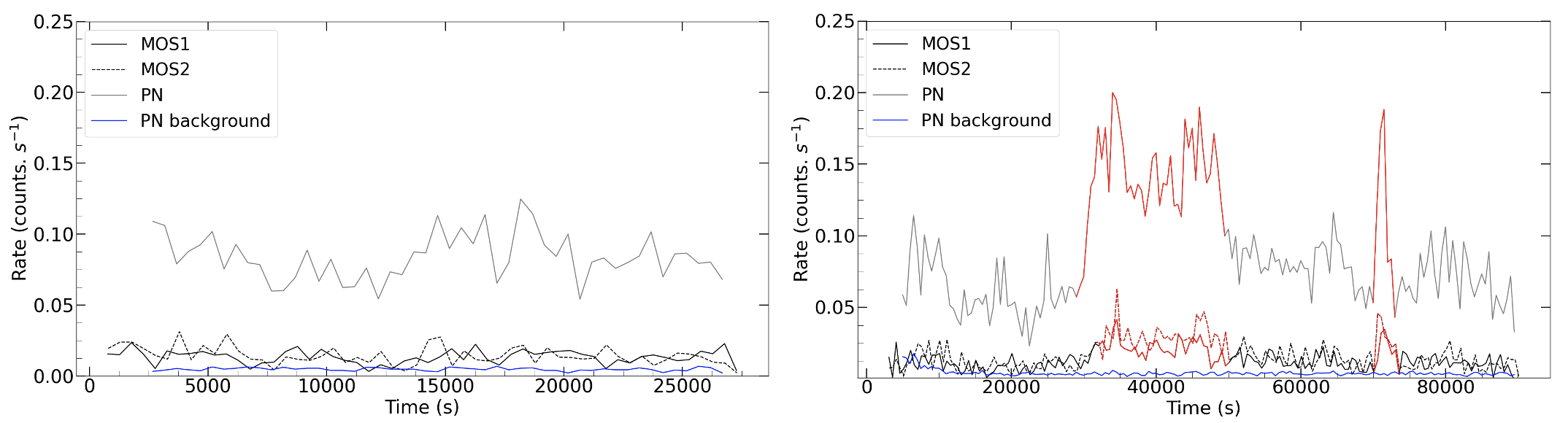}      
     \caption{The XMM–Newton X-ray light curves of HD~223889, with 500\,s time binnning in the energy range: 0.2-2\,keV. On the left side we show the signal from the two MOS detectors and the PN signal for "observation~1", while on the right side we show the signals from "observation~2".~In both panels we show the PN-background and background-subtracted light curves. The gray line represents the light curve of the PN~detector, while the solid-black line and black dashed line correspond to MOS1 and MOS2, respectively.~The PN-background is represented in blue.~The red lines in the right panel represent the parts of the observation where we identified flares by visual inspection based on the displayed elevated count rates.} 
\label{LC}
\end{figure*}

The extracted X-ray images from the three CCD detectors are shown in Fig~\ref{fig:X-ray}.
A circular extraction region with a radius of 20~arcsec was centered on the expected proper motion corrected position of HD~223889 during the epoch of each XMM–Newton observation (blue dashed-circle). In addition, a source-free background region with a radius of 60~arcsec was carefully defined (magenta dashed-circle). Lightcurves and CCD spectra were extracted for the MOS and PN detectors, along with RGS1 and 2, according to the procedure from the XMM-Newton SAS 12.12.1 user \href{https://xmm-tools.cosmos.esa.int/external/xmm_user_support/documentation/sas_usg/USG/}{handbook}. We used the same data processing methods for both observations.

\begin{table}[h]
    \centering
    \caption{\textit{XMM-Newton} data of HD~223889}
    \resizebox{\columnwidth}{!}{
        \begin{tabular}{ccccc}
        \hline
        \hline
             & ObsDate & ObsID & Exposure time (ks)  \\
        \hline
             Obs 1 & April 2020 & 0840844101 & 28 \\
             Obs 2 & May 2022   & 0900940101 & 100 \\
        \hline
        \end{tabular}
    }
    \label{tab:datainfo}
\end{table}

\section{Results} \label{sect: results}

\subsection{Temporal variability of HD~223889’s corona}
We have extracted the lightcurve from the source and background regions of the two MOS cameras and the PN~detector for both observations with 500\,s time binning in the energy range: 0.2-2\,keV.~The gray line represents the light curve of the PN~detector, while the black solid and dashed lines correspond to MOS1 and MOS2, respectively. The darker lines in the right panel represent the flare~phases.
In observation~2~(Fig.~\ref{LC},~right~panel) the corona of HD~223889 shows a clear variability (shown in red). The resulting lightcurve shows two significant flares, one between 30~ks and 50~ks and the other between 69~ks and 73~ks. Therefore, we divided the lightcurve into two phases: the ``quiet''~phase and the ``flare''~phase. This distinction was important because reconnection events associated with flares usually introduce new material into the star's corona, potentially altering its physical and chemical characteristics \citep{1987SoPh..113..249H}.~Conversely, there was no evidence of flare activity during "observation~1"~(Fig.~\ref{LC},~\textcolor{black}{left} panel). Therefore, we consider the spectra obtained during "observation~1" as representative of a quiescent phase.

\subsection{HD~223889 coronal properties from X-ray spectra}
  
\begin{figure}[hbt!] 
     \centering
      \includegraphics[width=\columnwidth]{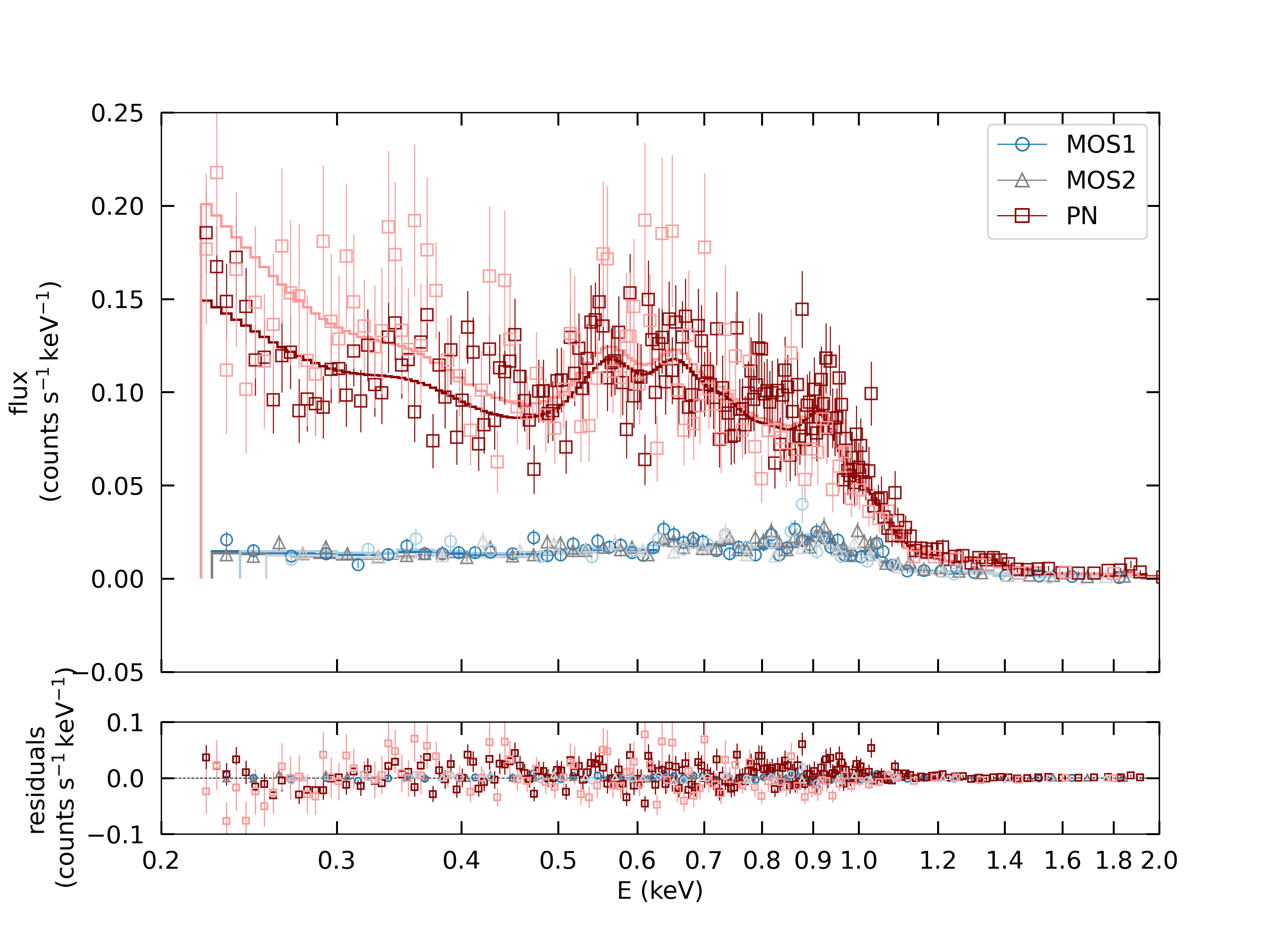}
         
     \caption{Spectra of the EPIC detector as a function of energy~(keV), with MOS1 presented by a blue circle, MOS2 by a gray triangle, and PN by red squares. The residuals of the fits are shown in the bottom row. The spectrum has a soft nature, with an average coronal temperature of 4.21~MK. Vivid colors correspond to "observation~2" quiescent phase, while weak colors represent "observation~1".}. 
\label{figure fitting spectra}
\end{figure}

\begin{figure}[hbt!] 
     \includegraphics[width=\columnwidth]{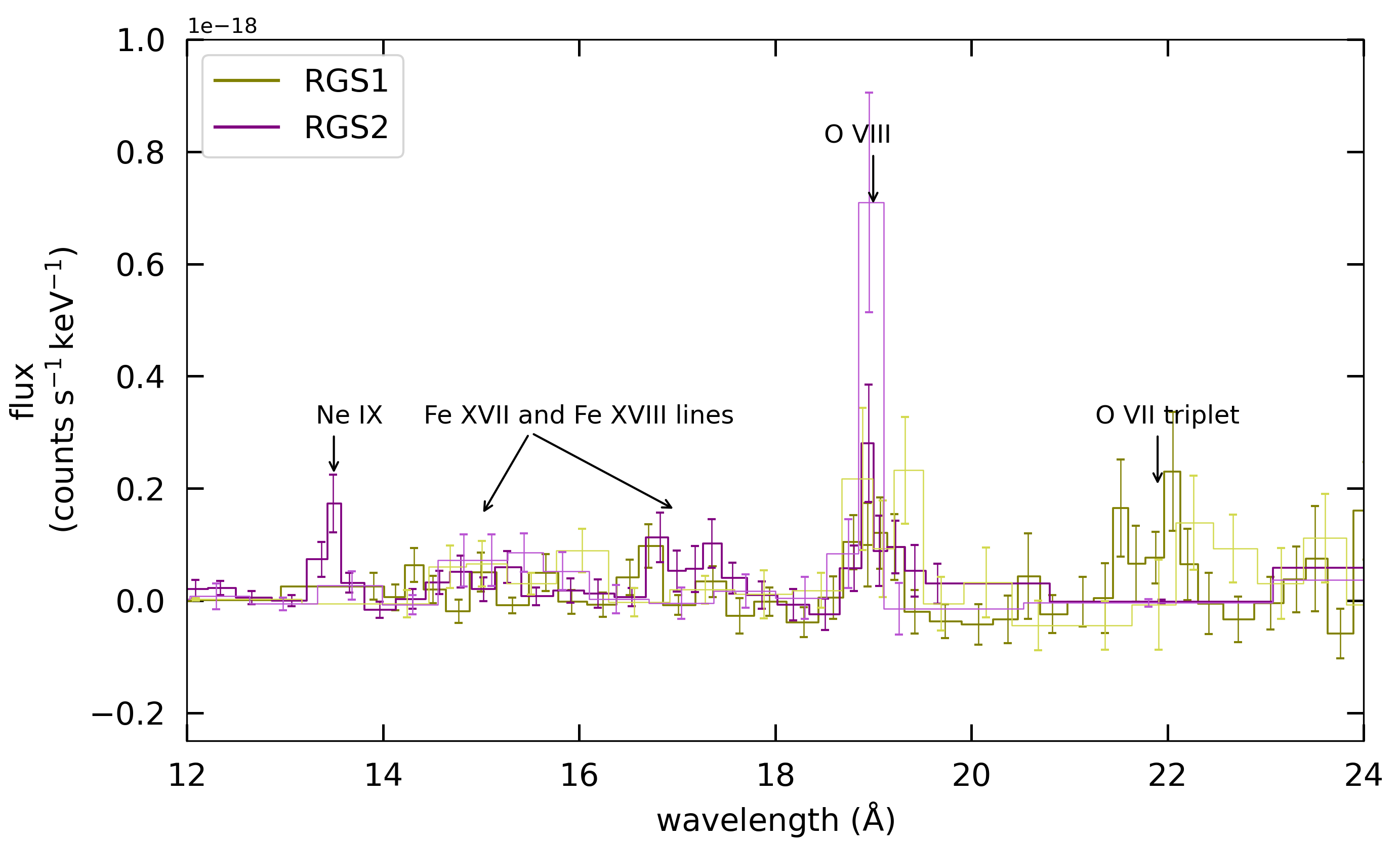}
     \caption{The spectra from the RGS instruments show flux~(counts\,$\rm s^{-1} keV^{-1}$) plotted against wavelength~($\AA$), with associated error bars. RGS1 is represented by a green line, and RGS2 by a purple line. Vivid colors indicate original "observation 2", while fainter colors represent "observation ". Key spectral lines, including NeIX, FeXVII, FeXVIII, OVIII, and the OVII triplet, are marked with arrows and annotations.}
     \label{fig: RGS}
\end{figure}

We extracted CCD~spectra of HD~223889 from two MOS~cameras, PN~detector, and RGS~spectra for both observations. For observation~2, we distinguished between two spectra for MOS1, MOS2, PN, RGS1, RGS2, one of which represents the quiescent phase and the other the original spectra with flaring. 
Using XSPEC version~12.12.1, we conducted a spectral fit with the coronal plasma model vAPEC \citep{2001ApJ...556L..91S} with the solar photospheric abundances from \cite{1998SSRv...85..161G}.

It is common to use the composition of the solar photosphere as a reference for M~dwarfs, considering how difficult it is to determine the abundances of these stars~\citep{2018ApJ...862...66W, 2012ApJ...753...76W, 2010ApJ...717.1279W}. This challenge arises from their low effective temperature, which leads to the formation of predominantly molecular lines in their spectra. 

The vAPEC model within XSPEC describes an optically thin thermal plasma, as appropriate for a stellar corona, where individual elemental abundances can be fitted\footnote{\url{http://www.atomdb.org/}}. A notable aspect of the vAPEC~model is the ability to account for variable abundances of different elements in the plasma. When using the vAPEC for spectrum fitting, several free parameters are available to optimize the fit of the model to the observed data. These parameters include the temperature~($kT$), which represents the thermal plasma temperature, usually measured in keV. The parameters also include the coronal abundances, which indicate the abundance values for different elements that can be varied independently.
In addition the vAPEC model fits each temperature component with its corresponding emission measure\footnote{Definitions can be found in the XSPEC documentation at \url{https://heasarc.gsfc.nasa.gov/docs/xanadu/xspec/manual/manual.html}}.

Figure~\ref{fig: RGS} shows the high-resolution spectra of RGS1 (green) and RGS2 (purple) for both observations. They are of rather low S/N, but the expected strong spectral line complexes are recognizable (Ne IX, Fe XVII, and Fe XVIII~lines, O VIII, and O VII as marked in the plot). We therefore treat oxygen (O), neon (Ne) and iron (Fe) as free parameters in our spectral fits, while the abundances of the other elements were fixed to a value of 1.

In Table~\ref{fitting_results}, we present the results of the spectral fitting.~We distinguish between two different spectra; quiescent and flare.~The quiescent spectra composed of "observation~1" and "observation~2" quiescent phase~(Obs$_{\rm 1}$ + Obs$_{2}^{\rm Q}$), while the flare spectra is defined by fitting the \textcolor{black}{flare~phases} in \textit{observation~2}~($\rm{Obs}_{2}^{\rm F}$).

We found that a coronal model with three temperature components was necessary to yield a satisfactory spectral fit.~Uncertainties for the fitted spectral model parameters were determined using XSPEC's \texttt{error}~command.~The latter adjusts the value of the desired parameter while keeping all other parameters fixed at their best-fit values.~The best-fit parameters are those that result in the lowest chi-square value, indicating the best match between the model and data.~The reported range correspond to a 1-$\sigma$ uncertainty.
For the quiescent spectra fitting,~our fit yields a satisfactory fit with a reduced chi-squared value of $\rm \chi_{red}^2$ =~1.2.; the flaring spectrum yielded a similarly satisfactory fit with $\rm \chi_{red}^2$~=~1.18.

In Fig.~\ref{figure fitting spectra}, we present the best-fit model of the quiescent spectra.~The top panel displays the MOS1, MOS2, and PN spectra, while the bottom panel shows the residuals of the fits.~We focus on the quiescent spectra because the results are comparable to those of the flare spectra, except for the flux, which is higher in the flare spectra, as shown in Table~\ref{fitting_results}.

The latter shows the parameters of the best-fit model for fitting~1 and 2, respectively. The table also shows the emission measure characterized by the parameter "norm" of the vAPEC model along with the X-ray flux in the energy band 0.2–2~keV.

The properties of HD 223889, summarized in Table~\ref{figure fitting spectra}, classify it as a moderately active M~dwarf based on its X-ray luminosity, which ranges from \( 2.98 \times 10^{27} \, \text{erg s}^{-1} \) during quiescence to \( 5.91 \times 10^{27} \, \text{erg s}^{-1} \) during flares. In comparison, low-activity M dwarfs have X-ray luminosities below \( 10^{27} \, \text{erg s}^{-1} \), while highly active M dwarfs exceed \( 10^{28} \, \text{erg s}^{-1} \).~The star also an X-ray surface flux of $\log F_{\rm X, surf}$~=~5.26, which is lower than any other M~dwarf that has been studied for iFIP patterns~(see Figure~4d in \citealt{2012ApJ...753...76W}).

We also calculate the average coronal temperature of HD~223889 for the case of quiescent and flaring spectra fitting to be respectively 4.21\,MK and 5.14\,MK. Coronal temperatures between 2 and 7 MK emit both soft and hard X-rays. At lower temperatures (2-4\,MK), soft X-rays (energies < 2 keV) dominate due to the plasma's peak emissivity in this region.

\begin{table}
\Large
\centering
\caption{\textcolor{black}{Best-fitting parameters of the coronal model for HD~223889 from MOS, PN, and RGS data.}}
\resizebox{\columnwidth}{!}{

\begin{tabular}{p{70mm} p{50mm} p{40mm}}
\hline
\hline
\textcolor{black}{\textbf{Parameter}} & \textcolor{black}{\textbf{Quiescent \newline(Obs$_{\rm 1}$ + Obs$_{2}^{\rm Q}$)}} & \textcolor{black}{\textbf{Flaring \newline(Obs$_{\rm 2}^{\rm F}$)}} \\[2pt]
\hline \\
\textcolor{black}{k$T_{\rm 1}$ (keV)} & \textcolor{black}{0.15$^{+0.02}_{-0.02}$} & \textcolor{black}{0.15$^{+0.03}_{-0.03}$} \\[5pt]

\textcolor{black}{k$T_{\rm 2}$ (keV)} & \textcolor{black}{0.32$^{+0.02}_{-0.02}$} & \textcolor{black}{0.34$^{+0.03}_{-0.03}$} \\[5pt]

\textcolor{black}{k$T_{\rm 3}$ (keV)} & \textcolor{black}{0.94$^{+0.05}_{-0.05}$} & \textcolor{black}{0.92$^{+0.10}_{-0.10}$} \\[5pt]

\textcolor{black}{EM$_{\rm 1}$ ($ \times 10^{\rm +49} \rm cm^{-3}$)} & \textcolor{black}{4.11$^{+0.49}_{-0.37}$} & \textcolor{black}{3.97$^{+1.20}_{-1.15}$} \\[5pt]

\textcolor{black}{EM$_{\rm 2}$ ($ \times 10^{\rm +49}\rm cm^{-3}$)} & \textcolor{black}{6.57$^{+0.61}_{-0.61}$} & \textcolor{black}{9.05$^{+2.20}_{-1.87}$} \\[5pt]

\textcolor{black}{EM$_{\rm 3}$ ($ \times 10^{\rm +49}\rm cm^{-3}$)} & \textcolor{black}{2.01$^{+0.24}_{-0.24}$} & \textcolor{black}{4.39$^{+0.94}_{-0.89}$} \\[5pt]

\textcolor{black}{O} & \textcolor{black}{0.64$^{+0.07}_{-0.06}$} & \textcolor{black}{0.70$^{+0.09}_{-0.07}$} \\[5pt]

\textcolor{black}{Ne} & \textcolor{black}{1.25$^{+0.11}_{-0.11}$} & \textcolor{black}{1.56$^{+0.15}_{-0.12}$} \\[5pt]

\textcolor{black}{Fe} & \textcolor{black}{0.54$^{+0.09}_{-0.07}$} & \textcolor{black}{0.46$^{+0.07}_{-0.06}$} \\[5pt]

\textcolor{black}{Flux (erg cm$^{\rm −2} \rm s^{\rm  -1}$), 0.2-2\,keV} & \textcolor{black}{$2.44^{+0.10}_{-0.14} \times 10^{-13}$} & \textcolor{black}{$4.84^{+0.10}_{-0.22}~\times 10^{-13}$} \\[5pt]

\textcolor{black}{$L_{\rm X}$\, (erg $ \rm s^{−1}$), 0.2-2\,keV} & \textcolor{black}{$2.98^{+0.13}_{-0.17}\times 10^{27}$} & \textcolor{black}{$5.91^{+0.11}_{-0.28}\times 10^{27}$} \\[5pt]

\textcolor{black}{$F_{\rm bias}$} & \textcolor{black}{$0.24^{+0.19}_{-0.19}$} & \textcolor{black}{$0.39^{+0.20}_{-0.20}$} \\[5pt]
\hline

\end{tabular}}
\tablefoot{Column 2 lists the values obtained from fitting the quiescent spectra~(Obs$_{\rm 1}$ + Obs$_{2}^{\rm Q}$), while column 3 presents the results from fitting the flare spectra~(Obs$_{\rm 2}^{\rm F}$).}

\label{fitting_results}

\end{table}

\subsection{FIP or iFIP~effect?}
Our main goal is to quantify the FIP or iFIP effect in the corona of HD~223889.
The metric $F_{\rm bias}$ has been defined in the literature to characterize the trend of the coronal abundances within a stellar corona; specifically,  the coronal abundance measurements of high-FIP elements compared chosen low-FIP elements are encapsulated in this dimensionless number.
To determine the $F_{\rm bias}$, we follow a similar methodology as outlined in \cite{2012ApJ...753...76W} and \cite{2010ApJ...717.1279W}, and shown in equation~\ref{FIPbias}.

\begin{equation}
\mathrm{F_{bias} = avg \Sigma \log[\text{A/Fe}]_{\rm corona} - \log[\text{A/Fe}]_{\rm photosphere}}
\label{FIPbias}
\end{equation}

However, in contrast to their approach, which uses individually resolved spectral lines, we use here the fitted coronal abundances of the elements with very strong coronal emission lines as shown in Table~\ref{fitting_results}, namely Ne and O as high-FIP elements and Fe as the low-FIP element. These are normalized by solar photospheric abundances to derive the $F_{\rm bias}$ value; no stellar photospheric abundances are available for our observed target.

In cases of a solar-like~FIP effect, the $F_{\rm bias}$ takes on a negative value, whereas a positive value indicates the presence of an inverse FIP (iFIP) effect. Our results for HD~223889 reveal an iFIP pattern in both the quiescent and flaring spectra, with $F_{\rm bias}$ values of approximately 0.24 and 0.39, respectively. The associated formal uncertainties were determined through Gaussian error propagation from the fitted coronal abundances, resulting in a symmetric error of around 0.04 for the quiescent $F_{\rm bias}$ and 0.06 for the flare.

The uncertainty in our $F_{\rm bias}$ value is quite small compared to more detailed analyses such as in \cite{2012ApJ...753...76W} which deal with individually resolved spectral lines and should therefore in principle contain more information than our approach. This may indicate that our low-resolution spectral analysis underestimates the uncertainties somewhat. We therefore looked at the scatter in coronal Fe/O ratios determined from low-resolution X-ray spectra versus high-resolution X-ray spectra as shown in \cite{2018ApJ...862...66W},~their~figure~4. By applying a bootstrap technique with 1000 iterations, we determined an overall uncertainty of 0.153 associated with low-resolution versus high-resolution coronal abundance ratio determinations. We therefore added this systematic uncertainty to our formal fitting error bars mentioned above, to arrive at the final uncertainties on the $F_{\rm bias}$ parameter that is listed in table~\ref{fitting_results}. In light of those uncertainties, the variation in $F_{\rm bias}$ between the flare and quiescent phases is not significant.

\section{Discussion} \label{sect: discussion}

In Fig.~\ref{Parameter_exploration} we show the $T_{\rm eff}$-$F_{\rm bias}$ diagram with the additional $F_{\rm bias}$ of HD~223889.~The diamond symbol represents the stars from~\cite{2018ApJ...862...66W} and the square symbol corresponds to the additional stars from the sample of \cite{2022A&A...659A...3S}.~The data points are color coded by there corresponding $\log L_{\rm X}$.~Each data point is plotted with its corresponding error bars for the effective temperature~($T_{\rm eff}$) and the FIP~bias value~($F_{\rm bias}$). The 2 gray lines represent one of the two fitted lines for each one of the trends seen in \cite{2022A&A...659A...3S} that corresponds to their Eq.~2 and 4.

The sample of stars from \cite{2018ApJ...862...66W} includes a mixture of single and binary main sequence stars representing a wide range of X-ray luminosities. range from $\log L_{\rm X} = 26.99$ to $\log L_{\rm X}=30.06$.~These values were measured directly from the LETGS spectra within the canonical \textit{ROSAT} PSPC bandpass for soft X-ray of 0.1-2.4\,keV (e.g., 5–120\,$\AA$).

To compare different stars over the same energy range, we converted the X-ray luminosity of HD~223889 from the energy range \( 0.2{-}2 \, \text{keV} \) range to \( 0.1{-}2.4 \, \text{keV} \) 
using the Xspec ''flux'' command.~By doing so we obtained $\log L_{\rm X}$~=~27.52 for the quiescent spectra and $\log L_{\rm X}$ =~27.85 for the flaring spectra in the 0.1{-}2.4 \, \text{keV} energy band, i.e.\ only marginally larger values as expected.~The energy adjustment ensures that our results remain consistent and comparable with other studies.

The study by \cite{2022A&A...659A...3S} aimed to expand the $T_{\rm eff}$–$F_{\rm bias}$ diagram, originally developed by \cite{2012ApJ...753...76W} and refined by \cite{2015LRSP...12....2L}, to include evolved stars. Previously focused on main-sequence stars like Sun-like stars and M dwarfs, the authors compiled data on active stars with known coronal abundances to calculate $F_{\rm bias}$. Their results indicated nearly parallel branches in the $T_{\rm eff}$–$F_{\rm bias}$ relation, with a notable extension for low-mass M dwarfs ($T_{\rm eff} < 4000$ K).~The bimodal distribution of $F_{\rm bias}$ values was hinted at in \cite{2018ApJ...862...66W},~their Fig.~7a.

\begin{figure}[hbt!] 
     \centering
         \includegraphics[width=\columnwidth]{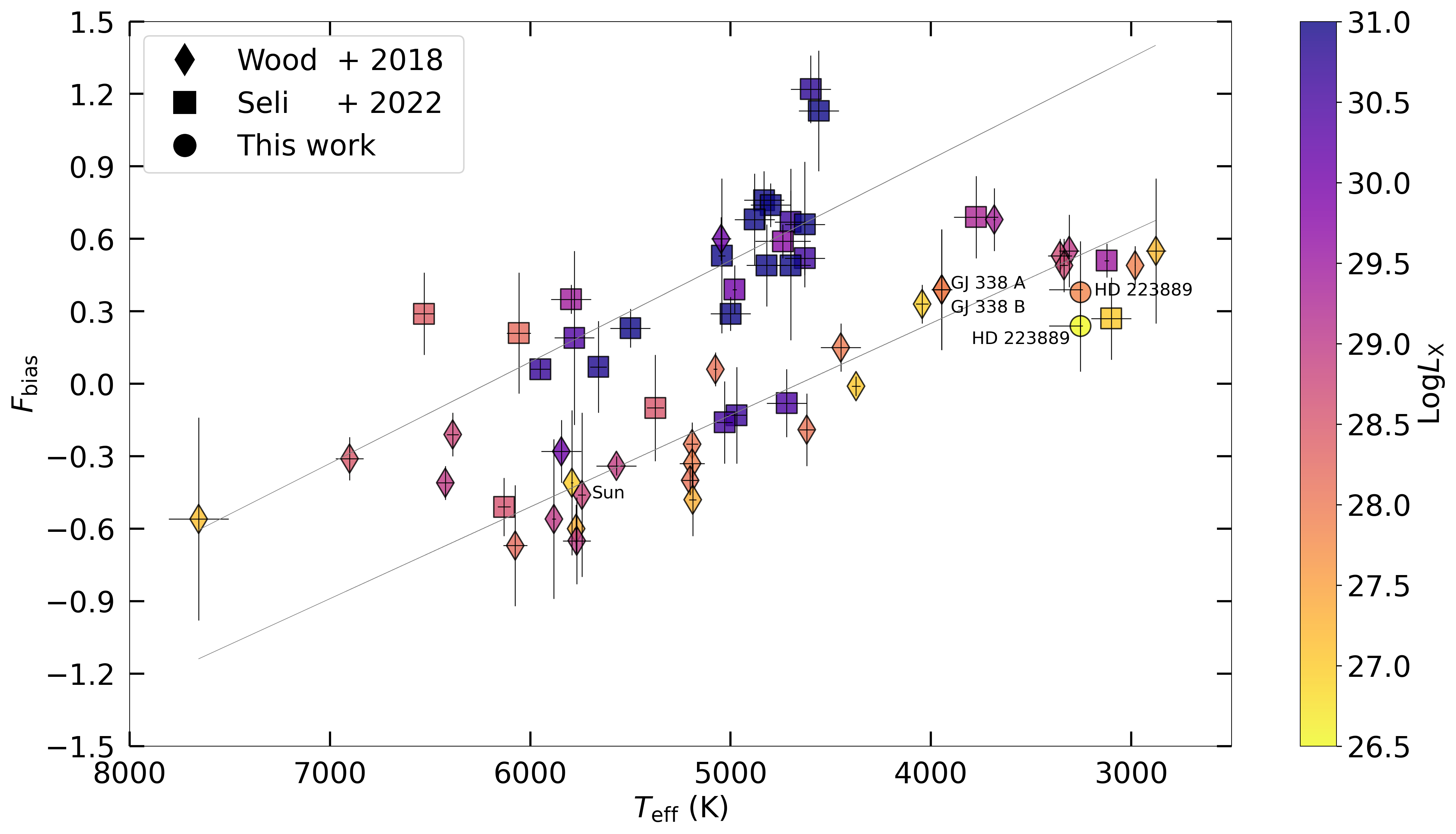}
     \caption{$T_{\rm eff}$-$F_{\rm bias}$ diagram. The diamond symbols represent the stars from \cite{2018ApJ...862...66W} and the square symbol corresponds to the additional stars from the sample of \cite{2022A&A...659A...3S}. The circles represent the $F_{\rm bias}$ of HD~223889 for the flare and the quiescent phase. Each data point is plotted with the corresponding error bars for the effective temperature~($T_{\rm eff}$) and the FIP~bias value~($F_{\rm bias}$). The 2 gray lines represent the fit of the lower and upper branches from \cite{2022A&A...659A...3S}, Eq.~2 and 4 respectively.~The color-coded data points indicating $\log~L_{\rm X}$~(X-ray luminosity) which is also taken as an activity indicator.}  
\label{Parameter_exploration}
\end{figure}

According to the $T_{\rm eff}$-$F_{\rm bias}$ relationship described in \cite{2010ApJ...717.1279W} and \cite{2018MNRAS.479.2351W}, an iFIP~effect would be expected for HD~223889 that is consistent with the pattern observed in M~dwarfs, regardless of their activity level.~This relationship is particularly evident in the almost linear progression of stellar composition across spectral types F to M in the X-ray spectra of moderately active stars ($L_{\rm X} <$ $10^{29}$\,erg s$^{-1}$).~From our low-resolution spectra we have \textcolor{black}{calculated} $F_{\rm bias}\sim 0.24$ for quiescent fitted spectra and $F_{\rm bias}$$\sim 0.39$ for the flare spectra with large uncertainties~for HD~223889~(Section \ref{FIPbias}).

It is worth noting that our study uses similar sample constraints as previous works such as \cite{2010ApJ...717.1279W}, i.e.\ we focus on stars with low to moderate X-ray luminosities. Studies including high-activity stars do find a larger fraction of stellar coronae displaying iFIP patterns than our sample. Observing a wide range of stellar activity levels in the M dwarf regime is challenging, since their small stellar radii mean that truly low activity M dwarfs typically do not produce enough X-ray photons to collect spectra of high enough quality for a coronal abundance study. This is also true for our target; however, we can still place it in context with other M dwarfs studied for their coronal abundances.

Our target HD~223389 can be compared to the wide binary star GJ~338. A detailed analysis of GJ~338~AB is given in \citealt{2018MNRAS.479.2351W}). The targets exhibit roughly similar characteristics, including radius, X-ray surface flux, coronal temperature, and activity levels, with GJ~338 AB being slightly more active than HD~223389.~Therefore it is not surprising that our calculated $F_{\rm bias}$ value of HD~223389 aligns closely with that of GJ~338~AB~($F_{\rm bias} = 0.39 \pm 0.25$, \citealt{2012ApJ...753...76W}).
This suggests that HD~223389 and the GJ~338~AB wide binary system may have similar coronal filling factors, indicating comparable fractions of active regions.~In other words, they have similarly constrained volumes for magnetic field expansion (\citealt{2019ApJ...875...35B}; \citealt{2021ApJ...909...17L}). Consequently, Alfvén waves generated in the chromosphere struggle to \textcolor{black}{expand} effectively, which increases their likelihood of reflecting and refracting back into the chromosphere.

The observations of HD~223889, alongside earlier findings on GJ~338~AB and other more active M~dwarfs, prompt an exploration into the possible existence of a plateau in the $T_{\rm eff}$–$F_{\rm bias}$ relationship for moderately active M~dwarfs; i.e., there may be a hint of a flattening of the slope in Fig.~\ref{Parameter_exploration} indicated by the data points.~With these results, the main question still remains open:
\textcolor{black}{Is there a minimum activity level for M~dwarfs below which the iFIP~pattern is no longer observed?}~Future X-ray missions which allow to collect X-ray spectra also for intrinsically X-ray fainter and therefore less active M~dwarfs, such as~NewAThena~\citep{2013arXiv1306.2307N, 2013arXiv1306.2333S} may shed significant light on this.

\section{Conclusions} \label{sect:conclusion}

In this study, we have performed \textcolor{black}{an analysis of the coronal} abundance patterns of
HD~223889 using XMM-Newton~data.~The analysis was limited to the modelling of MOS1, MOS2, PN, RGS1, RGS2, and uses a global 3~temperature VAPEC~model with three free parameters Ne, O, and Fe.~The presented fitting model in this study gave out the best result we can obtain given the low-resolution X-ray spectra.
This star is the lowest activity M~dwarf star studied so far with respect to the FIP~effect.~Our findings reveal an iFIP~effect that closely resembles that of another moderately active binary star, GJ~338~AB, with a comparable error margin. These results hint of a possible plateau in the iFIP effect within the moderately active star regime. This motivates further investigation of a small sample of M dwarfs with low coronal temperatures, ideally ranging between 2 MK and 4 MK. Such information will enhance our understanding of the patterns and underlying causes of (i)FIP effects and will help determine if this trend is disrupted in M dwarfs at lower coronal temperatures.~Rather than providing definitive conclusions, our study has raised additional questions and avenues for future research.

Finally, a comprehensive study of elemental abundances in the corona provides valuable insight into the composition of energetic particles and serves as a representative sample of coronal material. This research is of particular importance for planetary habitability, as these particles can affect surface chemistry and thus influence the prospects for a planet's long-term habitability over geological timescales.

\begin{acknowledgements}
          J.J.C.\ would like to thank the referee for their detailed feedback which improved the results and the analysis of this paper.~J.J.C. and K.P. acknowledge support from the German Leibniz-Gemeinschaft under project number P67/2018.
\end{acknowledgements}

\end{document}